# Hierarchical Neural Representation of Dreamed Objects Revealed by Brain Decoding with Deep Neural Network Features

**Tomoyasu Horikawa[1], Yukiyasu Kamitani[1,2]***

[1]Computational Neuroscience Laboratories, ATR, Kyoto, Japan
[2]Graduate School of Informatics, Kyoto University, Kyoto, Japan

**Correspondence:** Yukiyasu Kamitani: kamitani@i.kyoto-u.ac.jp

**Running title:** Hierarchical neural representation of dreamed objects

**Abstract**

Dreaming is generally thought to be generated by spontaneous brain activity during sleep with patterns common to waking experience. This view is supported by a recent study demonstrating that dreamed objects can be predicted from brain activity during sleep using statistical decoders trained with stimulus-induced brain activity. However, it remains unclear whether and how visual image features associated with dreamed objects are represented in the brain. In this study, we used a deep neural network (DNN) model for object recognition as a proxy for hierarchical visual feature representation, and DNN features for dreamed objects were analyzed with brain decoding of fMRI data collected during dreaming. The decoders were first trained with stimulus-induced brain activity labeled with the feature values of the stimulus image from multiple DNN layers. The decoders were then used to decode DNN features from the dream fMRI data, and the decoded features were compared with the averaged features of each object category calculated from a large-scale image database. We found that the feature values decoded from the dream fMRI data positively correlated with those associated with dreamed object categories at mid- to high-level DNN layers. Using the decoded features, the dreamed object category could be identified at above-chance levels by matching them to the averaged features for candidate categories. The results suggest that dreaming recruits hierarchical visual feature representations associated with objects, which may support phenomenal aspects of dream experience.

**Keywords: dream, brain decoding, deep neural networks, hierarchical neural representations, functional magnetic resonance imaging**

# 1. Introduction

Dreaming during sleep is a universal human experience and one that is often accompanied by highly realistic visual scenes spontaneously generated by the brain. The most striking characteristic of visual dreaming is its similarity to the visual experience during waking hours, and dreaming generally incorporates features that are typical of the waking experience, such as shapes, objects, and scenes. These phenomenological similarities are considered to be underlain by neural substrates common to both the awake and sleep states, and a number of studies have sought to address the neural commonalities and differences of these contrasting states by analyses of regional brain activations (Maquet et al., 1996; Braun et al., 1997; Braun et al., 1998; Maquet, 2000; Hong et al., 2009; Dresler et al., 2011), and brain activity patterns for specific visual contents (Horikawa et al., 2013).

A previous work investigated the commonality of neural representations of visual objects and scenes between perception and dreaming, and demonstrated that the dreamed objects/scenes could be predicted from brain activity patterns during sleep using statistical decoders trained to predict viewed object/scene categories (Horikawa et al., 2013). In this study, the authors used decoders trained to predict categorical labels of viewed objects and scenes, the labels of which were constructed from subjects' dream reports. They thereby demonstrated decoding of dream contents from brain activity patterns during sleep using stimulus-trained decoders. The decoders trained on brain activity patterns in higher visual cortex showed higher accuracy than those trained on brain activity patterns in lower visual cortex. Their results suggest that visual dream contents are represented by discriminative brain activity patterns similar to perception at least in higher visual areas.

While this study demonstrated accurate decoding of categorical information on dreamed objects from higher visual areas, it still remains unclear whether or how multiple levels of hierarchical visual features associated with dreamed objects are represented in the brain. Because brain decoding through multi-voxel pattern classification algorithms often obscures what made the labeled brain activity patterns discriminable, it is not clear what levels of visual information, including multiple levels of hierarchical visual features and semantics, enabled the successful decoding.

Several recent studies have addressed this issue by using explicit models of visual features, and investigated neural representations of visual contents (Kay et al., 2008; Khaligh-Razavi and Kriegeskorte, 2014; Naselaris et al., 2015; Jozwik et al., 2016; Horikawa and Kamitani, 2015). These studies used multiple levels of visual features, including Gabor filters and features extracted from hierarchical models, to represent visual images by patterns of visual features. They thereby established links between brain activity patterns and visual features or modeled the representational space of brain activity patterns using visual features to address how each visual feature is used to represent seen or imagined visual images.

Among a large number of visual features, hierarchical visual features, such as those from deep neural networks (Khaligh-Razavi and Kriegeskorte, 2014; Horikawa and Kamitani, 2015), would be especially suited to represent objects: They are hierarchical in the sense that higher-level features are composed of the outputs from the previous lower-level features. The reason for their suitability is that those visual features achieve varying levels of invariance to image differences through hierarchical processing, including differences in rotation, position, scale, and other attributes, which are often observed in images even within the same object categories, and acquire robust object-category-specific representations.

Our previous study (Horikawa and Kamitani, 2015) investigated neural representations of hierarchical visual features associated with seen and imagined objects by representing objects using patterns of visual features. In this study, we asked subjects to imagine visual images of presented object names and analyzed imagery-induced brain activity in combination with hierarchical visual features to observe how hierarchical visual feature representations are used during mental imagery. We first used the visual features derived from various computational models to represent an object by a vector of visual features, and then trained statistical regression models to decode feature vectors of viewed objects from brain activity patterns measured as subjects viewed images of objects (stimulus-trained decoder). The trained decoders were then used to decode visual features of seen and imagined objects, and the decoded feature vectors were used to identify the object categories. Our analyses showed that the stimulus-trained decoders better predicted the low/high-level visual features of seen objects from lower/higher visual areas respectively, showing a homology between the brain and deep neural networks. This provided empirical support for the idea that the deep neural networks can be a good proxy for the hierarchical visual system for object recognition. We further demonstrated high decoding accuracy for mid- to high-level features of imagined objects from relatively higher visual areas, suggesting the recruitment of feature-level representations during mental imagery, in particular for the mid- to high-level feature representation. Therefore, the same strategy would also be applicable to investigate hierarchical feature representations of dreamed objects, and may reveal the recruitment of feature-level representations associated with dreamed objects at least for mid- to high-level feature representations as the previous study demonstrated for volitional mental imagery (Horikawa and Kamitani, 2015).

Here, we investigated whether multiple levels of hierarchical visual features associated with dreamed objects are represented in the brain in a manner similar to perception. For this purpose, we applied the same strategy in Horikawa and Kamitani (2015) to the decoding of hierarchical visual features associated with dreamed objects from brain activity patterns during sleep. We used a deep convolutional neural network (DNN) for object recognition as a proxy for hierarchical visual feature representation. We represented images of objects using patterns of visual features derived from DNN models (Figures 1A, B). We then performed decoding analyses of DNN features associated with dreamed objects from brain activity patterns obtained from sleeping subjects. We used the decoders trained to decode visual features of seen objects, thereby testing whether visual dream contents are represented by the hierarchical feature representations elicited in visual perception. We also tested whether the decoded feature vector could be used to identify the reported dreamed object by matching it to the averaged feature vectors of images in multiple candidate categories. The results were then compared with those for seen and imagined objects (Horikawa and Kamitani, 2015) to see the differences in hierarchical representation and the ability to decode arbitrary objects beyond training categories ('generic object decoding').

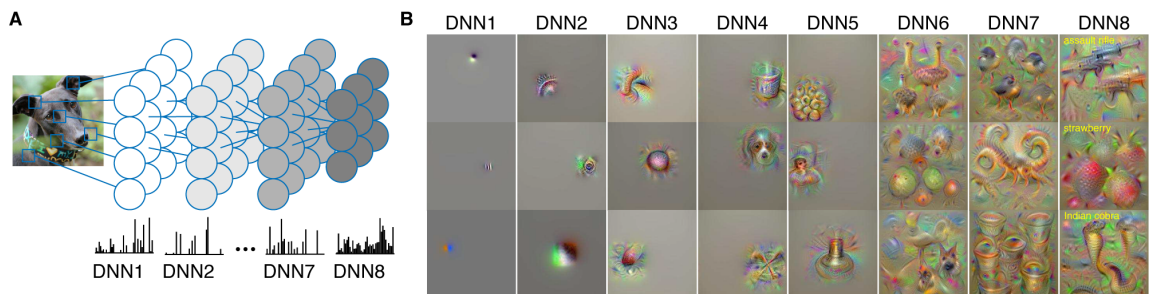

**Figure 1 | Deep neural network features.** (**A**) Representing an input image by visual feature patterns derived by DNN. The DNN was used to extract feature values of individual units in each layer. (**B**) Preferred images for each DNN layer. The images that highly activate each DNN unit were generated using activation maximization methods (see Materials and Methods: "2.8. Synthesis of preferred images using activation maximization" for details).

## 2. Materials and Methods

The data used for this study came from two previous studies performed at our laboratory (Horikawa et al., 2013; Horikawa and Kamitani, 2015). In these studies, two subjects (Subjects 1 and 2 in this study) participated in both of the two studies as Participants 1 and 3 in Horikawa et al. (2013) and as Subjects 1 and 2 in Horikawa and Kamitani (2015). Here, we provide a brief description of the subjects, datasets, and preprocessing of the MRI data for the main experiments. For full details, see Horikawa et al. (2013) and Horikawa and Kamitani (2015).

### 2.1. Subjects

Two healthy subjects (males, aged 27 and 42) with normal or corrected-to-normal vision participated in the experiments. Both subjects had considerable experience of participating in fMRI experiments, and were highly trained. Both subjects provided written informed consent for their participation in the experiments, in accordance with the Declaration of Helsinki, and the study protocol was approved by the Ethics Committee of ATR. The experimental of each subject were collected over multiple scanning sessions spanning over 2 years.

### 2.2. Dataset from Horikawa et al. (2013) ("dream" dataset)

We used an fMRI dataset from the sleep experiments conducted in a previous dream decoding study (Horikawa et al., 2013). This "dream" dataset was used for testing decoding models trained on part of the dataset from Horikawa and Kamitani (2015). A brief description of the dataset is given in the following paragraph (see Horikawa et al., 2013 for all experimental details).

fMRI signals were measured with simultaneous recording of electroencephalography (EEG) while subjects slept in an MRI scanner. They were awakened when a characteristic EEG signature was detected during sleep-onset periods (non-rapid eye movement [NREM] periods) and were then asked to provide a verbal report, freely describing their visual experience (NREM dream) before awakening. This procedure was repeated until at least 200 awakenings associated with a visual report were collected for each subject. From the collected reports, words describing visual objects or scenes were manually extracted and mapped to *WordNet*, a lexical database in which semantically similar words are grouped together as *synsets* (categories), in a hierarchical structure (Fellbaum, 1998). Using the semantic hierarchy, extracted visual words were grouped into *base synsets* that appeared in at least 10 reports from each subject (26 and 16 synsets for Subjects 1 and 2, respectively). The fMRI data obtained before each awakening were labeled with a *visual content vector*, each element of which indicated the presence or absence of a base synset in the subsequent report.

Note that these fMRI data were collected during sleep-onset periods (sleep stage 1 or 2) rather than rapid-eye movement (REM) periods. Although REM sleep and its underlying neurophysiological mechanisms were originally believed to be indispensable for dreaming, there has been accumulating evidence that dreaming is dissociable from REM sleep and can be experienced during NREM sleep periods (Nir and Tononi, 2009).

In addition to the fMRI data, we used visual images presented in the perception experiment described in Horikawa et al. (2013) to construct *category features* (see Materials and Methods: "2.7 Category feature vector") for dreamed objects (216 and 240 images for each category for Subjects 1 and 2, respectively).

## 2.3. Datasets from Horikawa and Kamitani (2015) ("training", "perception", and "imagery" datasets)

We used fMRI data from the perception experiment and the imagery experiment conducted in Horikawa and Kamitani (2015). The perception experiment had two sessions: a training image session and a testing image session. Data from the training image session of the perception experiment were used to train decoding models in this study ("training" dataset), which were then tested on the dream dataset from Horikawa et al., (2013). For comparison with the results from the dream dataset, the data from the perception experiment (the testing image session) and the imagery experiment in Horikawa et al. (2013) were also used as test datasets in this study ("perception" and "imagery" datasets). A description of the datasets is given in the following paragraph (see Horikawa and Kamitani, 2015 for all experimental details).

In the perception experiment of Horikawa et al. (2013), stimulus-induced fMRI signals were collected from two distinct sessions: a training image session and a testing image session, consisting of 24 and 35 separate runs respectively. Each run contained 55 stimulus blocks consisting of 50 blocks with different images and five randomly interspersed repetition blocks where the same image as in the previous block was presented. In each stimulus block an image (image size, 12 × 12 degree) was flashed at 2 Hz for 9 s. Images were presented on the center of the display with a central fixation spot. To indicate the onset of the block, the color of the fixation spot changed for 0.5 s before each stimulus block began. Subjects maintained steady fixation throughout each run, and performed a one-back repetition detection task on the images to maintain their attention on the presented images, responding with a button press for each repetition. In the training image session, a total of 1,200 images from 150 different object categories (eight images per each category) were each presented only once. In the testing image session, a total of 50 images from 50 object categories (one image from each category) were presented 35 times (blocks) each. Note that the categories in the testing image session were not used in the training image session. The presentation order of the categories was randomized across runs.

In the imagery experiment of Horikawa et al. (2013), the subjects were required to visually imagine images from one of the 50 object categories used in the testing image session of the perception experiment. The imagery experiment consisted of 20 separate runs, with each run containing 25 imagery blocks. Each imagery block consisted of a 3-s cue period, a 15-s imagery period, a 3-s evaluation period, and a 3-s rest period. During the rest periods, a fixation spot was presented in the center of the display. From 0.8 s before each cue period began, the color of the fixation spot changed for 0.5 s to indicate the onset of the blocks. During the cue period, words describing the names of the 50 categories presented in the testing image session of the perception experiment were visually presented around the center of the display (one target and 49 distractors). The word of the category to be imagined was presented with a red color (target), while the other words were presented in black (distractors). The onset and end of the imagery periods were signaled by beep sounds. Subjects were required to start imagining as many object images pertaining to the category described by the red word as possible. Their eyes were closed from the first beep sound to the second beep sound. After the second beep sound, the word of the target category was presented at the center of the display to allow the subjects to evaluate the vividness of their mental imagery on a five-point scale (very vivid, fairly vivid, rather vivid, not vivid, cannot recognize or forget the target) by a button press. The 25 categories in each run were pseudo-randomly selected from 50 categories such that the two consecutive runs contained all the 50 categories.

## 2.4. fMRI data preprocessing

The first 9 s of scans from each run were discarded to remove instability effects of the MRI scanner. The acquired fMRI data were subjected to three-dimensional motion correction using SPM5 (http://www.fil.ion.ucl.ac.uk/spm). The data were then coregistered to the within-session high-resolution anatomical image of the same slices used for EPI, and then subsequently to a whole-head high-resolution anatomical image common across the two studies. The coregistered data were then reinterpolated to 3 × 3 × 3 mm voxels.

For the dream data from Horikawa et al. (2013), we created data samples by first regressing out nuisance parameters, including a linear trend, and temporal components proportional to six motion parameters from the SPM5 motion correction procedure, from each voxel amplitude for each run, and the data were then *despiked* to reduce extreme values (beyond $\pm 3SD$ for each run). After that, voxel amplitudes around awakening were normalized relative to the mean amplitude during the period 60–90 s prior to each awakening. This period was used as the baseline, as it tended to show relatively stable blood oxygenation level dependent (BOLD) signals over time. The voxel values averaged across the three volumes (9 s) immediately before awakening served as a single data sample (the time window was shifted for time course analysis).

For the perception and imagery data from Horikawa and Kamitani (2015), we created data samples by first regressing out nuisance parameters, including a constant baseline, a linear trend, and temporal components proportional to six motion parameters from the SPM5 motion correction procedure, from each voxel amplitude for each run, and the data were then *despiked* to reduce extreme values (beyond $\pm 3SD$ for each run). The voxel amplitudes were then averaged within each 9-s stimulus block (three volumes) or 15-s imagery block (five volumes), after shifting the data by 3 s (one volume) to compensate for hemodynamic delays.

For testing decoding models with the dream dataset, the trials in which the last 15-s epoch before awakening was classified as *wake* were not used for the following analyses, and those classified as sleep stage 1 or 2 were used. We analyzed the dream fMRI data in two ways: single category-based analysis with averaged trials, and multiple category-based analysis with individual trials. In the single category-based analysis, fMRI samples were further averaged for the dream trials containing the same category while disregarding the other reported categories. Thus, one data sample is labeled only by a single category. This preprocessing yielded 26 and 16 averaged fMRI samples for Subjects 1 and 2, respectively (corresponding to the numbers of the base synsets). In the multiple category-based analysis, individual fMRI samples were labeled by multiple reported categories at each awakening.

For testing with the perception and imagery datasets, the blocks of the same category were averaged (35 and 10 blocks averaged for perception and imagery, respectively). This procedure yielded 50 averaged fMRI samples (corresponding to the 50 test categories) for each of the perception and imagery datasets in each subject.

**2.5. Region of interest (ROI) selection**

V1, V2, V3, and V4 were delineated by a standard retinotopy experiment (Engel et al., 1994; Sereno et al., 1995). The lateral occipital complex (LOC), the fusiform face area (FFA), and the parahippocampal place area (PPA) were identified using conventional functional localizers (Kourtzi and Kanwisher, 2000; Kanwisher et al., 1997; Epstein and Kanwisher, 1998). For the analysis of individual visual areas, the following numbers of voxels were identified for V1, V2, V3, V4, LOC, FFA, and PPA, respectively: 1054, 1079, 786, 763, 570, 614, and 369 voxels for Subject 1; 772, 958, 824, 545, 847, 438, and 317 voxels for Subject 2. A continuous region covering LOC, FFA, and PPA was manually delineated on the flattened cortical surfaces, and the region was defined as the *higher visual cortex* (HVC). Voxels from V1–V4 and HVC were combined to define the *visual cortex* (VC; 4794 and 4499 voxels for Subject 1 and 2, respectively). For full details on the experiments for localizing the regions of interest, see Horikawa et al. (2013) and Horikawa and Kamitani (2015).

**2.6. Visual features derived from deep convolutional neural network**

Using the deep convolutional neural network (DNN) proposed in a previous study (Krizhevsky et al., 2012), we computed visual features from the images used in the fMRI experiments, and also from images from an online image database[1] where images are grouped according to the hierarchy in *WordNet* (Fellbaum, 1998). We used the *MatConvNet* implementation of DNN[2], which was trained with images in ImageNet to classify 1,000 object categories. The DNN consisted of five convolutional layers (DNN1–5) and three fully connected layers (DNN6–8), with some of these layers containing a huge number of feature units (e.g., 290,400 units in DNN1). We randomly selected 1,000 units in each of the layers from one to seven to reduce the computational load while making sure that the selection was unbiased, and used all 1,000 units in the eighth layer. We represented each image by a vector of those units' outputs.

**2.7. Category feature vector**

We constructed *category feature vectors* to represent object categories using visual features in each DNN layers. We first computed visual feature vectors for all images of categories in the *ImageNet* database (50 test categories for Horikawa and Kamitani (2015), and 15,314 candidate categories; Deng et al., 2009) and for images used in the perception test experiment in Horikawa et al. (2013) (26 and 16 dreamed categories for Subjects 1 and 2, respectively). Using the computed feature vectors, category feature vectors were constructed for all categories by averaging the feature vectors of images belonging to the same category. These procedures were conducted for each DNN layer to construct feature representations of individual object categories (*single-category feature vectors*). In addition to that, we also constructed *multi-category feature vectors* to represent multiple object categories reported at each awakening in the dream dataset using features in each DNN layer. The multi-category feature vectors were constructed by averaging multiple single-category feature vectors annotated by reported categories at each awakening.

**2.8. Synthesis of preferred images using activation maximization**

We used the activation maximization method to generate preferred images for individual units in each layer of the DNN model (Simonyan et al., 2014; Yosinski et al., 2015; Mahendran and Vedaldi, 2016; Nguyen et al., 2016). Synthesis of preferred images starts from a random image and optimizes the input image to maximally activate a target DNN unit by iteratively calculating how the image should be changed via backpropagation. This analysis was implemented using custom software written in MATLAB based on the original Python code provided in blog posts[3].

---

[1] (ImageNet; Deng et al., 2009; http://www.image-net.org/; 2011 fall release)
[2] (http://www.vlfeat.org/matconvnet/)

**2.9. Visual feature decoding analysis**

We constructed multivoxel decoders to predict a visual feature vector of a seen object from fMRI activities in multiple ROIs in the training dataset (Horikawa and Kamitani, 2015) using a set of linear regression models. In this study, we used the sparse linear regression algorithm (SLR; Bishop, 2006), which can automatically select important voxels for decoding, by introducing sparsity into weight estimation through Bayesian parameters estimation with the automatic relevance determination (ARD) prior (see Horikawa and Kamitani, 2015 for detailed descriptions). The decoders were trained to predict the values of individual elements in the feature vector (consisting of 1,000 randomly selected units for DNN1–7 and all 1,000 units for DNN8) using the training dataset (1,200 samples from the perception experiment).

Decoding accuracy was evaluated by the correlation coefficient between the category feature and decoded feature values of each feature unit. The correlation coefficients were pooled across the units and the subjects for each DNN layer and ROI.

These analyses were performed for each combination of DNN layers (DNN1–8) and brain regions of interest (V1, V2, V3, V4, LOC, FFA, and PPA), and the entire visual cortex covering all of the visual subareas listed above (VC). We performed voxel selection prior to the training of the regression model for each feature unit: voxels showing the highest correlation coefficients with the target variable (feature value) in the training data were used (at most 500 voxels for V1, V2, V3, V4, LOC, FFA, and PPA; 1,000 voxels for VC). For details of the general procedure of feature decoding, see Horikawa and Kamitani (2015).

**2.10. Pairwise identification analysis**

In the pairwise identification analysis, the category of a seen/imagined/dreamed object was identified between true and false categories, using the feature vector decoded from the averaged fMRI activity pattern for each object category. The decoded feature vector was compared with two candidate category feature vectors, one for the true category and the other for a false category selected from the 15,314 candidates. The category with a higher correlation coefficient was selected as the identified category. The analysis was repeated for all combinations of the test categories (50 categories for the perception and imagery datasets; 26 and 16 categories for the dream dataset of Subjects 1 and 2, respectively) and the 15,314 candidate categories. The accuracy for each test category was evaluated by the ratio of correct identification. This was further averaged across categories and subjects to characterize the accuracies with the dream, perception, and imagery datasets.

---

[3] (Mordvintsev, A., Olah, C., Tyka, M., DeepDream - a code example for visualizing Neural Networks, https://github.com/google/deepdream, 2015; Øygard, A.M.,Visualizing GoogLeNet Classes, https://github.com/auduno/deepdraw, 2015)

**2.11. Data and code availability**

The experimental data and codes used in the present study are available from the corresponding author upon request.

## 3. Results

We applied the decoders trained with stimulus-induced fMRI signals (stimulus-trained decoders) to dream dataset to test whether multiple levels of visual features associated with dreamed object categories could be decoded from brain activity of sleeping subjects. For this analysis, the dream fMRI samples (three-volume [9 s] averaged fMRI signals immediately before awakening) annotated by each individual object category were averaged across awakenings, and these averaged fMRI samples were used as input to the stimulus-trained decoders (Figure 2A). To evaluate the prediction accuracy in each unit, Pearson's correlation coefficient was calculated between the decoded and the single-category feature values for the series of test samples for each subject. The correlation coefficients were then averaged across all feature units obtained from two subjects for each DNN layer. Here, we used the correlation coefficient between the series of the category feature values and the decoded feature values in each unit, instead of the correlation coefficient between the category feature vector and the decoded feature vector for each sample. This was because we constructed decoders for each unit independently, and the baseline amplitude pattern across units alone could lead to spuriously high correlation coefficients between feature vectors.

The correlation coefficients between the features decoded from the dream fMRI dataset and the category features in multiple ROIs are shown in Figure 2B. While the absolute values of the correlation coefficients from dream fMRI dataset were lower than those from perception and imagery fMRI datasets (Horikawa and Kamitani, 2015), positive correlation coefficients were observed from decoders trained with relatively higher visual areas at mid- to high-level DNN layers (46 out of 56 pairs of DNN layers and ROIs, one-sided $t$ test after Fisher's Z transform, uncorrected $p < 0.01$). For most of the DNN layers, the previous study (Horikawa and Kamitani, 2015) showed that the decoding accuracy of seen category features was moderately high from most of visual areas with peak accuracy in V4, whereas the decoding accuracy of imagined category features was high for mid- to higher visual areas. The category feature decoding of dreamed objects showed highest correlation coefficients around the higher visual areas, suggesting the qualitatively similar tendency to the results for imagined rather than seen object categories (Horikawa and Kamitani, 2015).

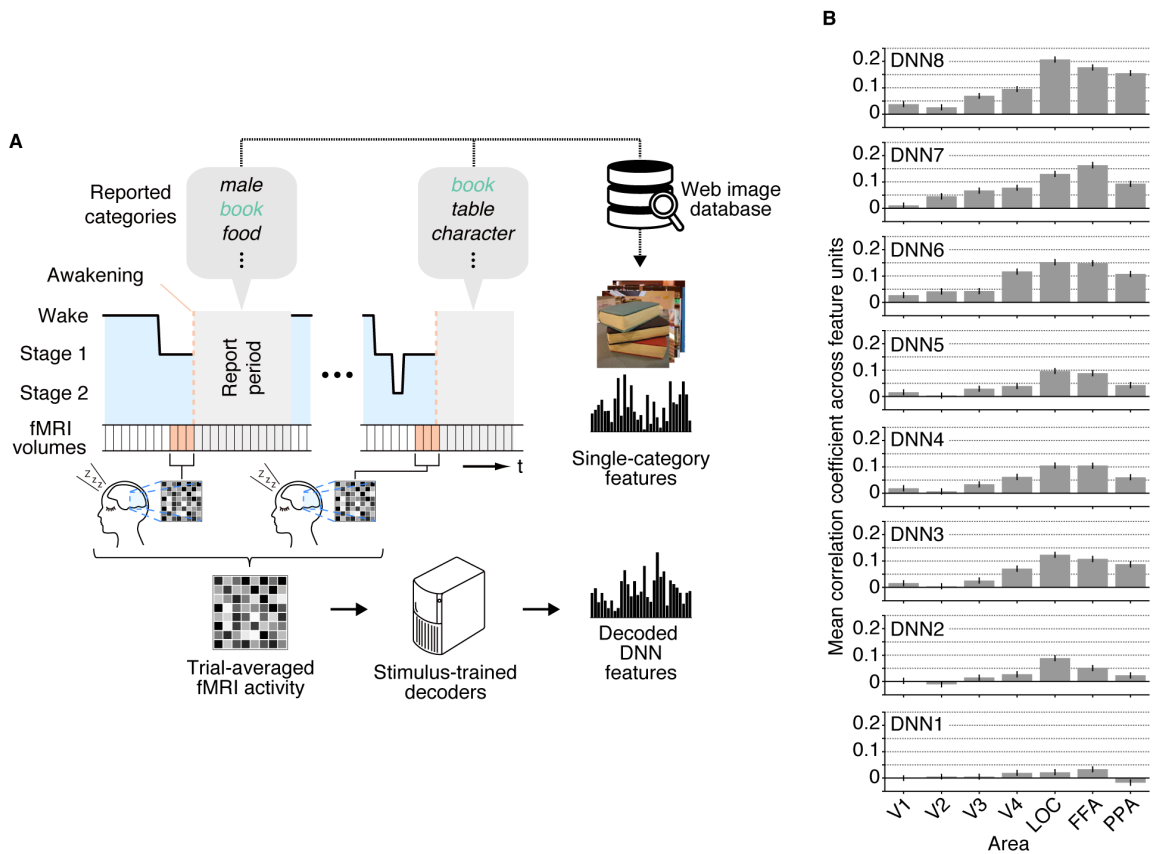

**Figure 2 | Single category feature decoding with averaged trials.** (**A**) A schematic view of analysis procedure. The decoders trained to decode the DNN feature values calculated from the presented images in the training dataset were applied to an averaged fMRI activity associated with a specific dream category to test decodability of the values of the category features constructed from multiple images annotated by the dream categories. (**B**) Correlation coefficients between the category feature values and the decoded feature values (error bars indicate 95% CI across feature units; two subjects pooled; three-volume [9 s] averaged fMRI signals immediately before awakening).

While we first focused on individual dream categories at a time by averaging fMRI samples at multiple awakenings annotated by the common dreamed object categories (Figure 2), we also performed decoding analysis on brain activity patterns from each awakening annotated by multiple dreamed object categories (Figure 3A). For this analysis, the same decoders were applied to fMRI samples at each single awakening (three-volume [9 s] averaged fMRI signals immediately before awakening) to obtain decoded features for all awakenings (samples classified as sleep stage 1 or 2). Then, the multi-category features were constructed by averaging the single-category features for multiple object categories reported at subsequent awakenings. The accuracy was evaluated by Pearson's correlation coefficients between decoded feature values and feature values of the multi-category features for the series of test samples. This analysis showed that feature values decoded from brain activity patterns in higher visual areas just before awakening positively correlated with feature values of the multi-category features constructed for object categories reported at subsequent awakening at mid- to high-level DNN layers (Figure 3B; 45 out of 56 pairs of DNN layers and ROIs, one-sided *t* test after Fisher's Z transform, uncorrected $p < 0.01$). The results suggest that single trial-based fMRI signals contain sufficient information to decode feature-level representations about dreamed object categories while the accuracy was relatively low.

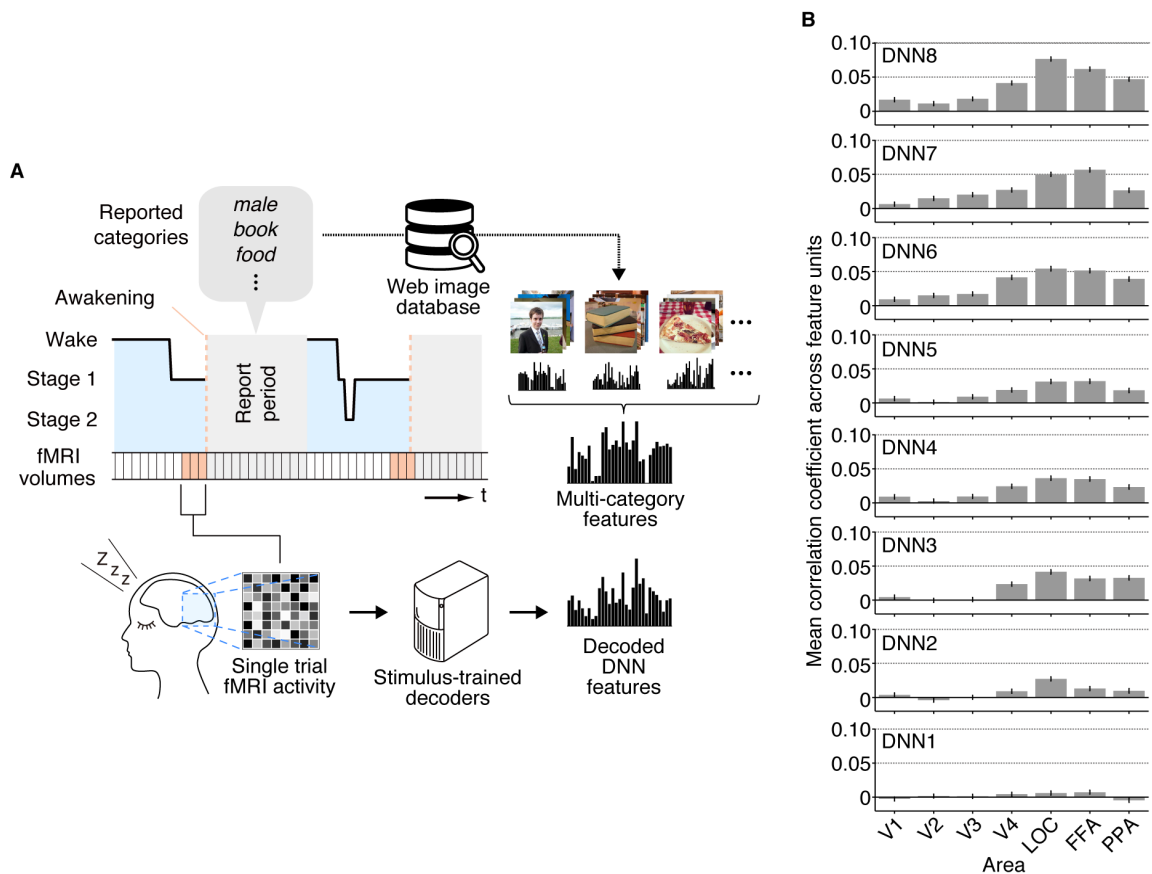

**Figure 3 | Multi-category feature decoding with individual trials**. (**A**) A schematic view of analysis procedure. The stimulus-trained decoders were applied to fMRI samples at each single awakening (or trial) in the dream fMRI dataset (orange areas, three-volume [9 s] averaged fMRI signals immediately before awakening). The values of the single-category features for reported dreamed objects were averaged to construct the multi-category features for each awakening. The accuracy was evaluated by Pearson's correlation coefficients between feature values of the decoded and the multi-category features for the series of test samples. (**B**) Correlation coefficients between multi-category features and decoded features (error bars, 95% CI across feature units; two subjects pooled).

Furthermore, when the time window for the feature decoding analysis was shifted around the time of awakening, the correlation coefficient peaked around 0 to 10 s before awakening for most of the DNN layers in both of the averaged- and single-trial analyses (Figure 4; no correction for hemodynamic delay). This is consistent with the results of the category decoding reported in the previous study (Horikawa et al., 2013). While the high correlations after awakening may be explained by hemodynamic delay and the large time window, the general tendency for the high correlations to be relatively prolonged, especially for higher DNN layers, may reflect feature representations associated with retrieved dream contents during reporting.

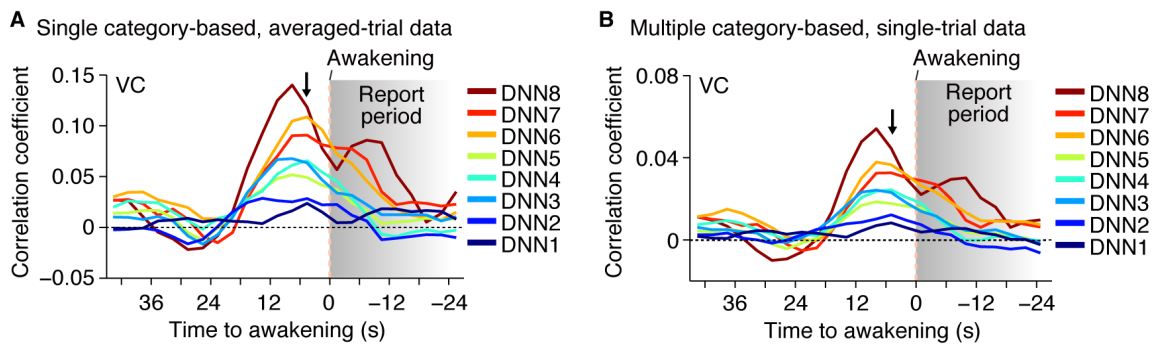

**Figure 4 | Time course of feature prediction**. Correlation coefficients were calculated between the decoded and the category feature values for the series of test samples (two subjects pooled; decoded from VC). The plot shows the mean feature prediction accuracy with the 9-s (three-volume) time window centered at each point (arrowed point for main analyses; cf. Figure 2B and 3B). (**A**) Time course of feature prediction with single category-based, averaged-trial data. (**B**) Time course of feature prediction with multiple category-based, single-trial data.

Finally, we tested whether the decoded feature vectors can be used to identify dreamed object categories and compared the results with the identification accuracies of seen and imagined object categories reported in a previous study (Horikawa and Kamitani, 2015). We did this by matching the decoded feature vectors and category feature vectors calculated from multiple images of candidate categories in the image database (Figure 5A). The pairwise identification accuracy for all combinations of the DNN layers and ROIs are shown in Figure 5B. For most combinations of the DNN layers and ROIs, the dreamed object categories can be identified from brain activity patterns with a statistically significant level (43 out of 56 pairs of layers and ROIs, one-sided $t$ test, uncorrected $p < 0.05$). Additionally, the analysis showed significantly high identification accuracy of dreamed objects from the LOC and FFA for all of the DNN layers (one-sided $t$ test, uncorrected $p < 0.05$). The pairwise identification accuracies for dreamed, seen, and imagined objects obtained by decoders trained on brain activity pattern in an entire visual cortex are shown in Figure 5C. The identification of dreamed objects showed a higher than chance accuracy for most of the DNN layers (one-sided $t$ test, uncorrected $p < 0.05$, except for DNN7). The identification accuracy for seen and imagined object categories was higher than the chance level for all of the DNN layers (one-sided $t$ test, uncorrected $p < 0.05$; re-analyzed using the datasets from Horikawa and Kamitani, 2015), with the highest accuracy shown around the mid-level layers. Similar to the results of the perception and imagery, the identification of the dreamed object categories also showed relatively higher accuracy at mid-level DNN layers around DNN5. However, the accuracy tendency across layers under the dream condition was slightly different from those under the perception and imagery conditions, in the sense that the highest layer, DNN8, also showed higher accuracy, whereas the DNN7 showed poor performance, which may suggest the unique characteristic of dream representations.

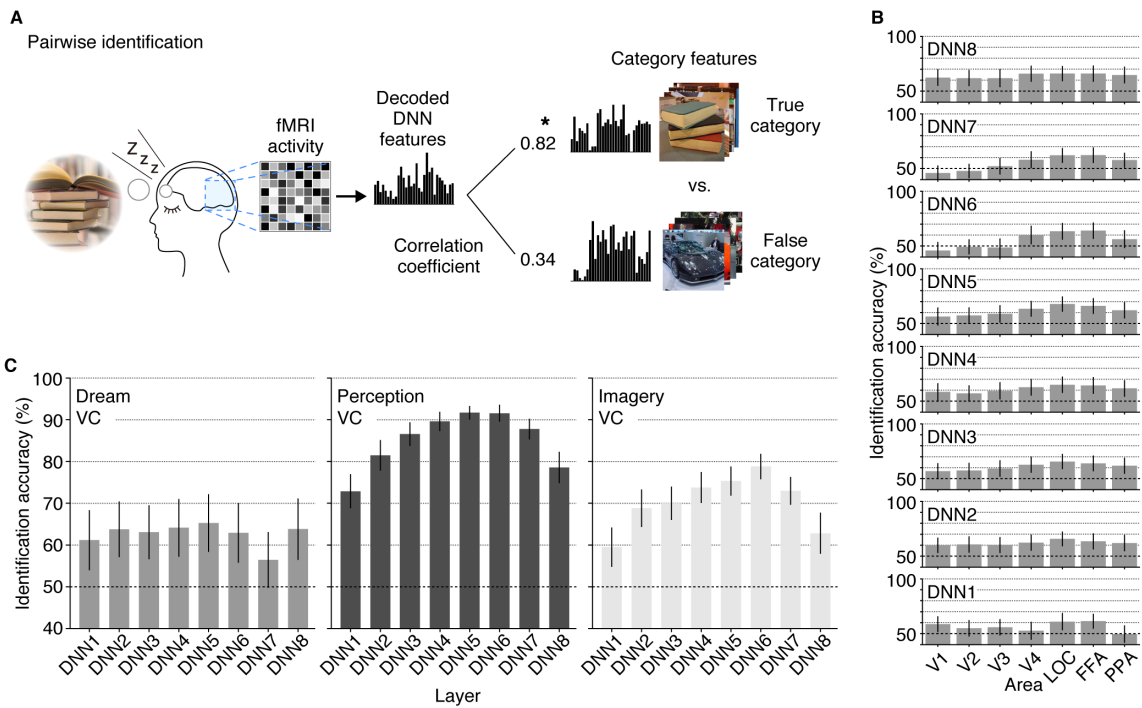

**Figure 5 | Pairwise identification analysis.** (**A**) A schematic view of the pairwise identification analysis. Given a decoded feature vector, the correlation coefficient was calculated between the decoded feature vector and the category feature vectors of two candidate categories (one true, the other false), and the category with a higher correlation coefficient was selected as the predicted category (star). (**B**) Pairwise identification accuracy of dreamed object categories for each combination of DNN layers and ROIs. The pairwise identification analysis was carried out for all combinations of DNN layers and ROIs. (**C**) Pairwise identification accuracy for the dream (left), perception (middle), and imagery (right) datasets. The visual feature vectors decoded from VC activity were used for identification analyses (error bars, 95% CI across samples; two subjects pooled; three-volume [9 s] averaged fMRI signals immediately before awakening).

## 4. Discussion

In this study, we examined whether hierarchical visual feature representations common to perception are recruited to represent dreamed objects in the brain. We used the decoders trained to decode visual features of seen object images and showed that the feature values decoded from brain activity during dreaming positively correlated with feature values associated with dreamed object categories at mid- to high-level DNN layers. This made it possible to discriminate object categories in dreaming at above-chance levels. These results reveal the recruitment of hierarchical visual feature representations shared with perception during dreaming.

In our analyses, we have shown that the multiple-levels of DNN features associated with dreamed objects can be predicted using the stimulus-trained decoders especially from the relatively higher visual areas (Figure 2B and 3B), as in our previous finding with imagined objects (Horikawa and Kamitani, 2015). The present results demonstrated not only semantic or categorical representations but also feature-level representations were recruited during dreaming to represent dreamed objects in a manner similar to perception. While a previous study demonstrated decoding of category information on dreamed objects (Horikawa et al., 2013), it did not clarify whether multiple levels of hierarchical visual feature representations are used to represent dream contents. In contrast to that, the present study demonstrated decoding of hierarchical visual features associated with dreamed object categories (Figure 2B and 3B), especially for features in mid- to high-level DNN layers (see Figure 1B for the characteristics of feature units in these layers), providing empirical evidence for recruitments of hierarchical visual feature representations during dreaming. These results further our understanding of how dreamed objects are represented in our brain.

Our decoding of feature-level representations of dreamed objects allowed us to discriminate dreamed object categories, although the accuracy is limited, without pre-specifying target categories, and achieved predictions beyond the categories used for decoder training: this characteristic was conceptualized as 'generic object decoding' in our previous brain decoding study (Horikawa and Kamitani, 2015). This framework is also known as the "*zero-data learning*" or "*zero-shot learning*" (Larochelle et al., 2008) in the machine-learning field, in which a model must generalize to classes with no training data. In the previous dream decoding study (Horikawa et al., 2013), the target categories to be decoded from brain activity patterns were determined from and restricted to the reported dream contents consisted of around twenty object categories for each subject. By contrast, we did decoding of dreamed object categories via predictions of deep neural network features. Thereby, we were able to decode arbitrary categories once the decoders were trained, even though the fMRI data for decoder training were collected irrespective of the reported dream contents. Our results extended the previous results on generic decoding of seen and imagined objects (Horikawa and Kamitani, 2015) to dreamed objects, demonstrating the generalizability of the generic decoding approach across different visual experiences.

The present study extended the previous results reporting recruitments of hierarchical visual feature representations during volitional mental imagery (Horikawa and Kamitani, 2015) to spontaneous mental imagery, in the sense that the feature-level representations were recruited without volitional attempt to visualize images. Taken together with the generalizability of the generic decoding from task-induced brain activity to spontaneous brain activity, we may be able to expect that the generic decoding approach via visual feature prediction is also applicable to decoding of visual information from other types of spontaneously generated subjective experiences, such as mind wandering (Smallwood and Schooler, 2015) or visual hallucination induced by psychedelic drugs (Carhart-Harris et al., 2016), which may help to understand the general principles of neural representations of our visual experience.

Our demonstration of the generic decoding of dreamed object categories indicates the commonality of hierarchical neural representations between perception and dreaming, but there still may be a representational difference between, perception, imagery and dreaming as suggested from different tendency in identification accuracy across DNN layers (Figure 5C). In our analyses, the identification accuracy of seen and imagined objects showed a single peak at around mid-level DNN layers (DNN4–7). On the other hand, the identification accuracy of dreamed objects showed poor accuracy at DNN7, whereas DNN8, which showed relatively poor accuracy for the seen and imagined conditions, showed higher accuracy (Figure 5C). Additionally, the high-level ROIs (LOC and FFA) rather than the mid-level ROI (V4) tended to show higher dreamed object identification accuracy for most of DNN layers (Figure 5B), while the identification of seen and imagined objects showed highest accuracy from V4 activity (Horikawa and Kamitani, 2015). Because there were differences in test categories between the perception/imagery datasets and dream dataset (50 test categories for perception and imagery; ~20 reported dream categories for dreaming) such difference may partially affect the results. However, the discontinuous profile of identification accuracy across DNN layers, relatively high accuracy in mid- and top-level DNN layers (DNN5 and 8) and low accuracy in DNN6 and 7, might suggest the involvement of higher cognitive functions, such as memory and abstract knowledge, to generate object representations during dreaming. Our time course analyses of the feature prediction accuracy showed the prolonged high accuracy for the high-level DNN features during reporting periods (Figure 4), which may also be explained by higher level cognitive functions related with memory retrieval and verbal reporting. The associations between dreaming and higher cognitive functions (Nir and Tononi, 2009) may lead to robust representations that resemble the high-level DNN layer.

While our analyses showed higher decodability for features at mid- to high-level DNN layers from relatively higher ROIs (Figure 2B, 3B), we were not able to provide evidence on how low-level features of dreamed objects are represented in lower ROIs. This was partly because our analyses were restricted to feature representations associated with object categories. Specifically, while the decoders were trained to decode visual features of individual images, the decoding accuracy was evaluated by correlation coefficients between the decoded features and the category features. Furthermore, the decoding from dream fMRI data was based on category-averaged brain activity and not based on brain activity induced by a specific image. Because of these limitations, our analyses should have reduced the sensitivity to the information on low-level image features. Thus, the poor accuracy for low-level DNN features and ROIs does not necessarily mean that we should reject the possibility of the recruitment of low-level image features in the representation of dream contents. Whether low-level image features, such as color and contrast, are represented in the dreaming brain is a topic that is worth addressing in future study.

**Conflict of Interest**

The authors declare that the research was conducted in the absence of any commercial or financial relationships that could be construed as a potential conflict of interest.

**Author contributions**

TH and YK designed the study. TH performed experiments and analyses. TH and YK wrote the paper.

**Funding**

This research was supported by grants from JSPS KAKENHI Grant number JP26119536, JP26870935, JP15H05920, JP15H05710, ImPACT Program of Council for Science, Technology and Innovation (Cabinet Office, Government of Japan), and the New Energy and Industrial Technology Development Organization (NEDO).

**Acknowledgements**

The authors thank Kei Majima and Mitsuaki Tsukamoto for helpful comments on the manuscript.